# Understanding different efficiency droop behaviors in InGaN-based near-UV, blue and green light-emitting diodes through differential carrier lifetime measurements

Lai Wang, Xiao Meng, Jiaxing Wang, Zhibiao Hao, Yi Luo, Changzheng Sun, Yanjun Han, Bing Xiong, Jian Wang, and Hongtao Li

Tsinghua National Laboratory for Information Science and Technology, Department of Electronic Engineering, Tsinghua University, Beijing 100084, China.

## Abstract

Efficiency droop effect under high injection in GaN-based light emitting diodes (LEDs) strongly depends on wavelength, which is still not well understood. In this paper, through differential carrier lifetime measurements on commercialized near-UV, blue, and green LEDs, their different efficiency droop behaviors are attributed to different carrier lifetimes, which are prolonged as wavelength increases. This relationship between carrier lifetime and indium composition of InGaN quantum well is believed owing to the polarization-induced quantum confinement Stark effect. Long carrier lifetime not only increases the probability of carrier leakage, but also results in high carrier concentration in quantum well. In other words, under the same current density, the carrier concentration in active region in near-UV LED is the lowest while that in green one is the highest. If considering the efficiency droop depending on carrier concentration, the behaviors of LEDs with different wavelengths do not show any abnormality. The reason why the efficiency droop becomes more serious under lower temperature can be also explained by this model as well. Based on this result, the possible solutions to conquer efficiency droop are discussed. It seems that decreasing the carrier lifetime is a fundamental approach to solve the problem.

## Introduction

GaN-based light-emitting diodes (LEDs) have been widely used in liquid crystal display backlighting and general lighting in recent years [1-3]. However, their efficiency suffers a serious degradation under high current density injection, which is well known as "efficiency droop" [4-6]. This effect hinders the development of LEDs, for its limitation on LED performance under high power, which is crucial for LED lighting. Therefore, the efficiency droop has attracted worldwide research interests from both academia and industry to explore its origins and solutions [7-56]. Unfortunately, after almost 10-year investigation, it is still not completely understood and well conquered up to now. Even so, people have gradually focused its origins on nonradiative Auger recombination [7, 11, 13, 21, 31, 42, 43, 46-48, 51, 53] inside the active region and/or carrier leakage [9, 14-16, 18, 20, 28, 29, 30, 32-24, 36, 37, 41, 45, 55] to outside of the active region (carriers are not injected efficiently). Moreover, it is also found that wider quantum well (QW) thickness [8, 11, 12], semipolar or nonpolar LEDs [57-62], or lower-indium-composition in QW [5, 63, 64] (violet or near-UV LEDs) suffers weaker efficiency droop effect.

It should be pointed out that the absolute value of efficiency cannot be ignored when we discuss the droop degree. The normalized efficiency curve depending on injection current density is insufficient to validate conclusion, since the relative reduction of efficiency is usually indeed small in poor LED samples [10]. For the state-of-the-art LEDs, the efficiency droop strongly depends on the emission wavelength [5]. Commonly, blue and green LEDs achieve their peak efficiency below 5 A/cm$^2$, but the efficiency suffers almost half degradation under 100 A/cm$^2$. And green LEDs usually exhibit much more serious droop effect than blue ones. On the other hand, InGaN-based near-UV (365~400 nm) LEDs show a slower efficiency increase as current density raises and reach the maximum under around 10~20 A/cm$^2$. Moreover, their droop degree is much weaker, keeping around 60~70% of the peak value under 100 A/cm$^2$. This wavelength-dependent (or indium-composition-dependent) phenomenon is so attractive and beneficial to clarify the origins of efficiency droop thoroughly.

In this paper, we measure the differential carrier lifetimes of commercialized near-UV, blue and green LEDs depending on injection current, using the method reported in Ref. 55 previously. It is found that the near-UV and green LEDs possess the shortest and longest carrier lifetime, respectively. We attribute this result to the polarization field in InGaN QW, which increases with indium composition and leads to slow radiative recombination rate. Similar to the assumption in Ref. 55, by ignoring the Auger recombination, the carrier injection efficiency and radiative recombination efficiency of these LEDs can be obtained by fitting the efficiency-current and carrier lifetime-current curves simultaneously. And then, the actual carrier concentrations in QWs can be calculated. It is found that the carrier concentration in near-UV LEDs is much lower than that in blue and green ones, even under the same current density. Both the longer carrier lifetime and higher carrier concentration in QW will result in the heavier carrier loss and hence the efficiency droop effect. Additionally, the solution for efficiency droop and the role of Auger recombination are also discussed.

### LEDs samples description

The commercialized LEDs measured in this paper are all grown on standard patterned sapphire substrate. The main structures of LEDs include an n-GaN bulk layer, n-InGaN/GaN superlattices, InGaN/GaN multi-quantum-well (MQW), a p-AlGaN electron blocking layer and a p-GaN contact layer. Their peak wavelengths are around 380, 460 and 530 nm, respectively, which are realized by adjusting the indium composition in InGaN QW. The MQW numbers of these samples are around 12-14, while the InGaN QW width keeps around 2.5~3 nm, resulting in the total active region thickness of 42, 35, 30 nm in near-UV, blue, and green LEDs, respectively. The chip size of blue and green LEDs are both 300×300 μm$^2$, while that of near-UV LED is 280×406 μm$^2$. And they are all packaged using resin lenses to enhance light extraction. The injection current-dependent light power, peak wavelength and forward voltage are measured in calibrated integral-spheres for visible LEDs and UV LED, respectively. Thus, the external quantum efficiency (EQE, $\eta_{ex}$) of LEDs at different injection current density can be calculated. In our analysis, the light extraction efficiencies (LEE, $\eta_{extr}$) of near-UV, blue and green LEDs are assumed constant of 50%, 80% and 80%, respectively. The lower $\eta_{extr}$ in the near-UV LED is owing to the absorption of UV light by the thick n-GaN bulk layer and also by ITO transparent electrode and resin lens. If the n-GaN bulk layer is replaced by an n-AlGaN bulk layer, the $\eta_{extr}$ will be increased. The differential carrier lifetimes of LEDs are measured using the method described elsewhere.[55]

### Results and Discussion

Figures 1(a) and 1(b) show the experimental internal quantum efficiencies (IQE, $\eta_{int}$) and carrier lifetimes ($\tau$) of near-UV, blue and green LEDs depending on injection current density ($J$), respectively, wherein the IQE $\eta_{int}$ is calculated as $\eta_{ex}/\eta_{extr}$. It should be mentioned that the IQE of the near-UV LED seems a bit lower than the state of the art. This is because the IQE of near-UV LED is more sensitive to the dislocation density[65, 66]. If instead of GaN buffer layer a sputtered AlN buffer layer is applied to lower the dislocation density[67], the IQE can be further improved to approach or even exceed that of blue one. However, it doesn't change the variation trend of IQE and the conclusion in following discussion. The $\eta_{int}$ vs. $J$ curves shown in Fig. 1(a) are identical to the typical droop behaviors of near-UV, blue and green LEDs, which are usually expressed by normalized plots[5]. As the current density increases, the carrier lifetimes' decreases are rapid first, and then turn slow gradually. And obviously the green LED possesses the longest carrier lifetime, while the near-UV one exhibits the shortest. The total carrier lifetime is determined by both radiative recombination lifetime and nonradiative recombination lifetime. The former is influenced by the piezoelectric polarization field in InGaN/GaN MQW[68], which would be stronger when indium composition is higher[69]. The polarization filed tilts the energy band of QW and separates the wave-functions of electron and hole, resulting in wavelength redshift and radiative recombination rate decay[70]. This is so-called the quantum confined Stark effect (QCSE). Due to its highest indium composition, the QCSE in green LED is the most serious among three samples, which is responsible for its longest radiative recombination lifetime [68]. On the other hand, the nonradiative recombination lifetime, which is close related to defects, would be much shorter in near-UV LED than that in blue and green ones. This is because the influence of dislocations on nonradiative recombination becomes more remarkable in near-UV LED due to its lowest carrier localization degree. These reasons lead to the differences of carrier lifetimes in near-UV, blue, and green LEDs shown in Fig. 1(b).

Both the IQE $\eta_{int}$ and carrier lifetime $\tau$ strongly depend on the carrier concentration $n$ in active region. According to the improved *ABC* model taking carrier delocalization effect into account [36, 55], the IQE and carrier lifetime can be expressed as [55]

$$\eta_{int} = \eta_{inj}\eta_{rad} = \eta_{inj}\frac{Bn^2}{An+Bn^2+Cn^3} \tag{1}$$

$$\frac{1}{\tau} = A + 2Bn + 3Cn^2 \tag{2}$$

$$A = r_L \cdot A_L + (1-r_L)A_{NL} \tag{3}$$

$$B = r_L \cdot B_L + (1-r_L)B_{NL} \tag{4}$$

$$r_L = \frac{1+\exp(-\frac{1}{k})}{1+\exp(\frac{n-n_C}{k \cdot n_C})} \tag{5}$$

wherein $\eta_{inj}$ is the current injection efficiency (CIE) and $\eta_{rad}$ is the radiative recombination efficiency (RRE). *A*, *B*, and *C* represent Schockley-Read-Hall (SRH) recombination, radiative recombination, and Auger recombination coefficients, respectively. Equations (3)-(5) reflect the carrier delocalization process as carrier concentration increases, which has been proven very important especially when current density is small. $A_L$, $B_L$ and $A_{NL}$, $B_{NL}$ represent the SRH and radiative recombination coefficients in localized centers and non-localized centers, respectively. $n_C$ is the density of localized centers and $k$ reflects the state density in these localized centers (a larger $k$ means a larger size and/or a deeper energy of localized center) [36, 55]. The carrier concentration $n$ is determined by

$$n = \frac{\eta_{inj}J\tau}{qd} \tag{6}$$

wherein $q$ is the electron charge and $d$ is the active region thickness.

According the equations above, we can obtain the unknown parameters by fitting the $\eta_{int}$ - *J* and $\tau$ – *J* curves simultaneously. In Ref. 55, it has been shown that if assuming $\eta_{inj}$=1 and the efficiency droop is induced only by Auger recombination, no convergent results can be achieved. Thus, in order to make model simple, Auger recombination is ignored temporarily in following analysis. However, we will still discuss its influence later. Assuming Auger recombination coefficient *C*=0, the fitting curves coincide exactly with the experimental data, as shown in Fig. 1(a) and (b). Meanwhile, $\eta_{inj}$ under different current density (shown in Fig. 2(a)) and the other parameters (shown in Table I) can be also obtained from the satisfactory fitting results.

In Table I, we can compare the carrier localization degrees in LEDs with different indium compositions. In previous studies, there are many evidences showing that indium composition fluctuation is the most probable reason for carrier localization and the fluctuation becomes stronger when the indium composition increases [65, 70, 71]. According to the fitting results, as wavelength of LED changes from near-UV to green, the parameters $n_C$ and $k$ increase from $1.35 \times 10^{17}$ cm$^{-3}$ and 2 to $4.8 \times 10^{17}$ cm$^{-3}$ and 4.4, respectively. As mentioned earlier, $n_C$ and $k$ reflect the carrier localization degree, thus these fitting results are in accord with the well-known published conclusion, which also indicates the rationality of the model and the fitting results. Based on the parameters in Table I, it is easy to calculate the RRE shown in Fig. 2(b). All the samples exhibit that RRE increases with the current density, which is a merited result when assuming Auger recombination coefficient $C=0$. The blue LED possesses the highest RRE, which is a widely accepted fact. However, the RRE of the present near-UV LED is the lowest among three samples, as it is the most sensitive to the dislocation density.

In Fig. 2(a), the CIE of all three samples decrease with increased current density. This is also the main cause of the efficiency droop under the assumption above. Moreover, the CIE of the near-UV LED is the highest among three samples, while that of the green one is the lowest. This result seems a little difficult to understand. As the near-UV LED commonly has the shallowest QW, wherein the carrier confinement should be weakest, the carrier leakage in it should be the most serious among three samples. We will talk about this phenomenon later.

Equation (6) tells us that the carrier concentration $n$ in active region of LED is determined not only by the current density, but also by the CIE and the carrier lifetime. So we can calculate $n$ using the CIE in Fig. 2(a) and the carrier lifetime $\tau$ in Fig. 1(b), and show its dependence on current density $J$ in Fig. 2(c). Although the CIE of the near-UV LED is the highest, its carrier concentration becomes the lowest due to the shortest carrier lifetime. Therefore, Fig. 2(c) exhibits the influence of carrier lifetime on carrier concentration clearly. This means the actual carrier densities in active region of different LEDs possibly vary considerably even though they operate under the same injection current density and their active region thicknesses are similar. Both long carrier lifetime and high carrier density are not beneficial for good luminescence. The former will increase the possibilities of carrier consumption by nonradiative recombination centers and carrier leakage from QW [45]. The latter not only leads to Auger recombination, but assists carrier leakage as well, for carriers can occupy higher energy level in QW.

In consideration of carrier concentration's importance, we can plot the IQE, carrier lifetime, CIE and RRE vs. carrier concentration curves, as shown in Fig. 3(a)~(d). The carrier lifetime and the RRE curves follow the similar trends with Fig. 1(b) and Fig. 2(b), respectively. However, interesting things appear in Fig. 3(a) and Fig. 3(c). In

Fig. 3(a), the efficiency droop behaviors in near-UV, blue and green LEDs become totally different. The droop slopes of blue and green LEDs seem almost the same when the IQE is plotted depending on carrier concentration, while in Fig. 1(a) green LED shows stronger efficiency droop effect. Moreover, the near-UV LED has the smallest droop no longer. Its droop degree seems close to or even a bit more serious than that of blue and green ones. Furthermore, in Fig. 3(c), the CIE curves are also replotted as a function of carrier concentration. Under the low carrier concentration, the near-UV LED holds the highest CIE, but unlike that in Fig. 2(a), it suffers the rapidest decline as carrier concentration increases. Compared with the green one, the blue LED actually is faced with the similar situation, which shows the earlier CIE droop. As a result, under the high carrier concentration, the green LED has the highest CIE instead. Thus, the perplexing CIE picture in Fig. 2(a) becomes clear in Fig. 3(c). Actually, the shallower the QW is, the more easily it leads to carrier leakage. Besides the wavelength-dependent efficiency droop behaviors, the carrier lifetime and carrier concentration can also be used to explain another phenomenon that efficiency droop becomes more serious at low temperature [28, 38]. As the nonradiative recombination rate becomes slow, the carrier lifetime will be prolonged at low temperature. This will result in the increase of carrier concentration in active region and hence the lower efficiency even under the same injection current. In addition, there have been many publications on improving LED droop performance by a variety of methods. However, we need to re-examine the root cause of these methods: Are they really improving the LED efficiency under the same carrier concentration or just reducing the carrier concentration because of a shortened carrier lifetime?

Fig. 3(a) and Fig. 3(c) show us that the efficiency droop in InGaN-based LED is strongly carrier concentration dependent. In this sense, current-dependent efficiency droop should rather be carrier concentration-dependent efficiency droop, though the former is more direct-viewing for LED industry. However, from a carrier concentration perspective, the efficiency droop seems unable to be overcome, since high carrier concentration induced recombination loss and carrier leakage are both unavoidable fundamental physical processes. Certainly, if other luminescence process is introduced, for example, the stimulated radiative recombination, it is another story. That is why laser diode is believed more attractive than LED in high power lighting [49]. On the other hand, if we want to improve the efficiency current-droop performance of LED, decreasing the carrier concentration is the most direct method. Increasing the QW width is a simple way to meet this requirement [8, 11, 12], but it sometimes will bring about the material quality decay due to the accumulated strain. Especially in c-plane LEDs, increasing the QW width will simultaneously strengthen the QCSE, and hence raise the carrier lifetime. According to Equation (6) and Fig.2 (c), a long carrier lifetime will lead to the high carrier concentration. Therefore, shortening the carrier lifetime is a more thorough way to overcome the current-dependent efficiency droop, which has been demonstrated in near-UV LED. For blue or green LED, wherein the intrinsic polarization field is stronger, some band engineering techniques are desired to shorten the radiative recombination lifetime, for instance, using the staggered QW [72] or thin barrier coupled QWs [16]. It should be pointed out that decreasing the nonradiative recombination lifetime can also result in the shorter carrier lifetime, but it will harm the RRE.

The nonpolar and semipolar LEDs developed in recent years provide a fundamental solution for QCSE. That's why they always exhibit the excellent efficiency droop-less performance [57-62], but they still need more time to reduce cost. Additionally, introducing surface plasmons into LED is another possible way to shorten the carrier lifetime [17, 26, 44], because they provide an extra channel for carrier recombination. But in order to prove their practicability in LED industry, it still needs more efforts to eliminate their consequent negative influences on LED efficiency.

Although the above discussion is based on Auger recombination ignoring, it does not mean it is inexistent. However, as satisfactory results cannot be obtained on fitting $\eta_{int} - J$ and $\tau - J$ curves simultaneously when we keep $Cn^3$ and let $\eta_{inj}=1$, we consider the influence of $Cn^3$ is weaker than $\eta_{inj}$. Especially, the carrier lifetimes remain almost unchanged as current density increases gradually, which show us no significant influence of Auger recombination on carrier lifetime. If the Auger recombination $Cn^3$ exists but is not significant, the assumption in the model above is reasonable. However, we cannot exclude another possibility. If the Auger recombination in GaN-based LED is not the classical one but the defect-assisted Auger recombination [13, 31], it might be expressed as $Cn^2$. In this defect-assisted Auger recombination, an electron transits to a defect energy level in band gap and transfers the energy to another electron in conduction band. The latter electron can transit to a higher energy level in conduction band, and subsequently has great probability to leak from active region. This Auger-assisted leakage will finally result in the low CIE [19, 42], meanwhile its influence on carrier lifetime will differ from the classical Auger recombination. These successive defect-assisted Auger recombination and Auger-assisted leakage processes seem quite likely to occur in LEDs according to recent reports on Auger recombination study [48, 51]. If it is true, the influence of Auger recombination actually has been included in CIE. Thus, the main conclusion of this paper will be still valid.

## Conclusions

In summary, the different efficiency droop behaviors in near-UV, blue, and green LEDs have been investigated. Through measuring and fitting the differential carrier lifetime measurements depending on injection current density, it is found that the carrier concentration in active region is affected by carrier lifetime intensively. Under the same injection current density, the near-UV LED has the lowest carrier concentration while the green one possesses the highest. By plotting the IQE and CIE curves as a function of carrier concentration, it is found that the efficiency droop depends on carrier concentration clearly. Therefore, the different efficiency droop behaviors among near-UV, blue and green LEDs become easier to understand. In fact, the near-UV LED has the most serious efficiency droop when carrier concentration increases due to its shallowest QW or weakest carrier confinement. But thanks to the shortest carrier lifetime, carrier concentration in its QW is the lowest, which makes it exhibit the weakest efficiency droop instead. On the contrary, the green LED is just the reverse. Based on this conclusion, shortening the carrier lifetime and hence reducing the carrier concentration in active region are regarded as a fundamental approach to overcome the efficiency droop in LEDs.

References

[1] D. A. Steigerwald, J. C. Bhat, D. Collins, R. M. Fletcher, M. O. Holcomb, M. J. Ludowise, P. S. Martin, and S. L. Rudaz, Illumination with solid state lighting technology, IEEE Journal of Selected Topics in Quantum Electronics, 8 310 (2002).

[2] M. H. Crawford, LEDs for solid-state lighting: performance challenges and recent advances, *IEEE Journal of Selected Topics in Quantum Electronics*, **15** 1028 (2009).

[3] S. Pimputkar, J. S. Speck, S. P. DenBaars, and S. Nakamura, Prospects for LED lighting, Nature Photonics, **3** 180 (2009).

[4] J. Piprek, Efficiency droop in nitride-based light-emitting diodes, Phys. Status Solidi A **207** 2217 (2010).

[5] J. Cho, E. F. Schubert, J. K. Kim, Efficiency droop in light-emitting diodes: Challenges and countermeasures, *Laser Photon. Rev.* **7** 408 (2013).

[6] G. Verzellesi, D. Saguatti, M. Meneghini, F. Bertazzi, M. Goano, G. Meneghesso, and E. Zanoni, Efficiency droop in InGaN/GaN blue light-emitting diodes: Physical mechanisms and remedies, *J. Appl. Phys.* **114** 071101 (2013).

[7] Y. C. Shen, G. O. Mueller, S. Watanabe, N. F. Gardner, A. Munkholm, and M. R. Krames, Auger recombination in InGaN measured by photoluminescence, *Appl. Phys. Lett.* **91** 141101 (2007).

[8] Y. L. Li, Y. R. Huang, and Y. H. Lai, Efficiency droop behaviors of InGaN∕GaN multiple-quantum-well light-emitting diodes with varying quantum well thickness. Applied Physics Letters **91** 181113 (2007).

[9] M. H. Kim, M. F. Schubert, Q. Dai, J. K. Kim, E. F. Schubert, J. Piprek, and Y. Park, Origin of efficiency droop in GaN-based light-emitting diodes, *Applied Physics Letters* **91** 183507 (2007).

[10] M. F. Schubert, S. Chhajed, J. K. Kim, E. F. Schubert, D. D. Koleske, M. H. Crawford, S. R. Lee, A. J. Fischer, G. Thaler, and M. A. Banas, Effect of dislocation density on efficiency droop in GaInN/GaN light-emitting diodes, *Applied Physics Letters* **91** 231114 (2007).

[11] N. F. Gardner, G. O. Müller, Y. C. Shen, G. Chen, S. Watanabe, W. Götz, and M. R. Krames, Blue-emitting InGaN–GaN double-heterostructure light-emitting diodes reaching maximum quantum efficiency above 200A∕cm2. *Applied Physics Letters*, **91** 243506 (2007).

[12] L. Wang, J. Wang, H. Li, G. Xi, Y. Jiang, W. Zhao, Y. Han, and Y. Luo, Study on injection efficiency in InGaN/GaN multiple quantum wells blue light emitting diodes, *Applied Physics Express* **1** 021101 (2008).

[13] J. Hader, J. V. Moloney, B. Pasenow, S. W. Koch, M. Sabathil, N. Linder, and S. Lutgen, On the importance of radiative and Auger losses in GaN-based quantum wells. *Applied Physics Letters*, **92** 261103 (2008).

[14] M. F. Schubert, J. Xu, J. K. Kim, E. F. Schubert, M. H. Kim, S. Yoon, S. M. Lee, C. Sone, T. Sakong, and Y. Park, Polarization-matched GaInN∕AlGaInN multi-quantum-well light-emitting diodes with reduced efficiency droop, Appl. Phys. Lett. **93**, 041102 (2008).

[15] J. Xie, X. Ni, Q. Fan, R. Shimada, Ü. Özgür, and H. Morkoç, On the efficiency droop in InGaN multiple quantum well blue light emitting diodes and its reduction with p-doped quantum well barriers, Appl. Phys. Lett. **93** 121107 (2008).

[16] X. Ni, Q. Fan, R. Shimada, Ü. Özgür, and H. Morkoç, Reduction of efficiency droop in InGaN light emitting diodes by coupled quantum wells, Applied Physics Letters, **93** 171113 (2008).

[17] K. C. Shen, C. Y. Chen, H. L. Chen, C. F. Huang, Y. W. Kiang, C. C. Yang, and Y. J. Yang, Enhanced and partially polarized output of a light-emitting diode with its InGaN/GaN quantum well coupled with surface plasmons on a metal grating. *Applied Physics Letters*, **93** 231111 (2008).

[18] J. Xu, M. F. Schubert, A. N. Noemaun, D. Zhu, J. K. Kim, E. F. Schubert, M. H. Kim, H. J. Chung, S. Yoon, C. Sone, and Y. Park, Reduction in efficiency droop, forward voltage, ideality factor, and wavelength shift in polarization-matched GaInN/GaInN multi-quantum-well light-emitting diodes. *Appl. Phys. Lett.* **94** 011113 (2009).

[19] K. J. Vampola, M. Iza, S. Keller, S. P. DenBaars, and S. Nakamura, Measurement of electron overflow in 450 nm InGaN light-emitting diode structures. *Applied Physics Letters*, **94** 061116. (2009)

[20] S. H. Han, D. Y. Lee, S. J. Lee, C. Y. Cho, M. K. Kwon, S. P. Lee, and S. J. Park, Effect of electron blocking layer on efficiency droop in InGaN/GaN multiple quantum well light-emitting diodes, *Applied physics letters* **94** 231123 (2009).

[21] K. T. Delaney, P. Rinke, and C. G. V. Walle, Auger recombination rates in nitrides from first principles, *Appl. Phys. Lett.* **94** 191109 (2009).

[22] Y. K. Kuo, J. Y. Chang, M. C. Tsai, and S. H. Yen, Advantages of blue InGaN multiple-quantum well light-emitting diodes with InGaN barriers. *Applied Physics Letters*, **95** 011116 (2009).

[23] H. Y. Ryu, H. S. Kim, and J. I. Shim, Rate equation analysis of efficiency droop in InGaN light-emitting diodes. *Applied Physics Letters*, **95** 081114 (2009).

[24] A. David and M. J. Grundmann, Droop in InGaN light-emitting diodes: A differential carrier lifetime analysis. *Appl. Phys. Lett.* **96** 103504 (2010).

[25] J. Hader, J. V. Moloney, and S. W. Koch, Density-activated defect recombination as a possible explanation for the efficiency droop in GaN-based diodes, *Appl. Phys. Lett.* **96** 221106 (2010).

[26] C. F. Lu, C. H. Liao, C. Y. Chen, C. Hsieh, Y. W. Kiang, and C. C. Yang, Reduction in the efficiency droop effect of a light-emitting diode through surface plasmon coupling. Applied Physics Letters, **96** 261104 (2010)

[27] Q. Dai, Q. Shan, J. Wang, S. Chhajed, J. Cho, E. F. Schubert, M. H. Crawford, D. D. Koleske, M. H. Kim, and Y. Park, Carrier recombination mechanisms and efficiency droop in GaInN/GaN light-emitting diodes. *Appl. Phys. Lett.* **97** 133507 (2010)

[28] J. Wang, L. Wang, W. Zhao, Z. Hao, and Y. Luo, Understanding efficiency droop effect in InGaN/GaN multiple-quantum-well blue light-emitting diodes with different degree of carrier localization, *Appl. Phys. Lett.* **97** 201112 (2010).

[29] C. H. Wang, C. C. Ke, C. Y. Lee, S. P. Chang, W. T. Chang, J. C. Li, Z. Y. Li, H. C. Yang, H. C. Kuo, T. C. Lu, and S. C. Wang, Hole injection and efficiency droop improvement in InGaN/GaN light-emitting diodes by band-engineered electron blocking layer. *Applied Physics Letters* **97** 261103 (2010).

[30] Q. Dai, Q. Shan, J. Cho, E. F. Schubert, M. H. Crawford, D. D. Koleske, and Y. Park, On the symmetry of efficiency-versus-carrier-concentration curves in GaInN/GaN light-emitting diodes and relation to droop-causing mechanisms. *Applied Physics Letters* **98** 033506 (2011).

[31] E. Kioupakis, P. Rinke, K. T. Delaney, and C. G. Van de Walle, Indirect Auger recombination as a cause of efficiency droop in nitride light-emitting diodes, *Appl. Phys. Lett.* **98** 161107 (2011).

[32] C. H. Wang, S. P. Chang, P. H. Ku, J. C. Li, Y. P. Lan, C. C. Lin, H. C. Yang, H. C. Guo, T. C. Lu, S. C. Wang, and C. Y. Chang, Hole transport improvement in InGaN/GaN light-emitting diodes by graded-composition multiple quantum barriers. *Applied Physics Letters*, **99** 171106 (2011).

[33] Y. Y. Zhang, and Y. A. Yin, Performance enhancement of blue light-emitting diodes with a special designed AlGaN/GaN superlattice electron-blocking layer. *Applied physics letters*, **99** 221103 (2011).

[34] D. S. Meyaard, G. B. Lin, Q. Shan, J. Cho, E. F. Schubert, H. Shim, M. H. Kim, and C. Sone, Asymmetry of carrier transport leading to efficiency droop in GaInN based light-emitting diodes, *Appl. Phys. Lett.* **99** 251115 (2011).

[35] S. Hammersley, D. Watson-Parris, P. Dawson, M. J. Godfrey, T. J. Badcock, M. J. Kappers, C. McAleese, R. A. Oliver, and C. J. Humphreys, The consequences of high injected carrier densities on carrier localization and efficiency droop in InGaN/GaN quantum well structures, J. Appl. Phys. **111** 083512 (2012).

[36] J. Wang, L. Wang, L. Wang, Z. Hao, Y. Luo, A. Dempewolf, M. Müller, F. Bertram, and J. Christen, An improved carrier rate model to evaluate internal quantum efficiency and analyze efficiency droop origin of InGaN based light-emitting diodes, *J. Appl. Phys.* **112** 023107 (2012).

[37] B. J. Ahn, T. S. Kim, Y. Dong, M. T. Hong, J. H. Song, J. H. Song, H. K. Yuh, S. C. Choi, D. K. Bae and Y. Moon, Experimental determination of current spill-over and its effect on the efficiency droop in InGaN/GaN blue-light-emitting-diodes. *Applied Physics Letters*, **100** 031905 (2012).

[38] D. S. Meyaard, Q. Shan, J. Cho, E. F. Schubert, S. H. Han, M. H. Kim, C. Sone, S. J. Oh, and J. K. Kim, Temperature dependent efficiency droop in GaInN light-emitting diodes with different current densities. *Applied Physics Letters*, **100** 081106 (2012).

[39] H. Y. Ryu, D. S. Shin, and J. I. Shim, Analysis of efficiency droop in nitride light-emitting diodes by the reduced effective volume of InGaN active material, *Appl. Phys. Lett.* **100** 131109 (2012).

[40] D. S. Shin, D. P. Han, J. Y. Oh, and J. I. Shim, Study of droop phenomena in InGaN-based blue and green light-emitting diodes by temperature-dependent electroluminescence, *Appl. Phys. Lett.* **100** 153506 (2012).

[41] G. B. Lin, D. Meyaard, J. Cho, E. F. Schubert, H. Shim, and C. Sone, Analytic model for the efficiency droop in semiconductors with asymmetric carrier-transport properties based on drift-induced reduction of injection efficiency. *Applied Physics Letters*, **100** 161106 (2012).

[42] M. Deppner, F. Römer, and B. Witzigmann, Auger carrier leakage in III-nitride quantum-well light emitting

diodes, *Phys. Status Solidi RRL* **6** 418 (2012).

[43] R. Vaxenburg, E. Lifshitz, and Al. L. Efros, Suppression of Auger-stimulated efficiency droop in nitride-based light emitting diodes, *Applied Physics Letters* **102** 031120 (2013).

[44] H. S. Chen, C. F. Chen, Y. Kuo, W. H. Chou, C. H. Shen, Y. L. Jung, and C. C. Yang, Surface plasmon coupled light-emitting diode with metal protrusions into p-GaN. *Applied Physics Letters*, **102** 041108 (2013).

[45] D. S. Meyaard, G. B. Lin, J. Cho, E. F. Schubert, H. Shim, S. H. Han, M. H. Kim, C. Sone, and Y. S. Kim, Identifying the cause of the efficiency droop in GaInN light-emitting diodes by correlating the onset of high injection with the onset of the efficiency droop. *Applied Physics Letters*, **102** 251114 (2013)

[46] M. Binder, A. Nirschl, R. Zeisel, T. Hager, H. J. Lugauer, M. Sabathil, D. Bougeard, J. Wagner, and B. Galler, Identification of nnp and npp Auger recombination as significant contributor to the efficiency droop in (GaIn) N quantum wells by visualization of hot carriers in photoluminescence, *Appl. Phys. Lett.* **103** 071108 (2013).

[47] F. Bertazzi, X. Zhou, M. Goano, G. Ghione, and E. Bellotti, Auger recombination in InGaN/GaN quantum wells: A full-Brillouin-zone study, *Applied Physics Letters* **103**, 081106 (2013).

[48] J. Iveland, L. Martinelli, J. Peretti, J. S. Speck, and C. Weisbuch, Direct measurement of Auger electrons emitted from a semiconductor light-emitting diode under electrical injection: identification of the dominant mechanism for efficiency droop, *Physical Review Letters*, **110** 177406 (2013).

[49] J. J. Wierer, J. Y. Tsao, and D. S. Sizov, Comparison between blue lasers and light-emitting diodes for future solid−state lighting. *Laser & Photonics Reviews*, **7** 963 (2013).

[50] D. P. Han, D. G. Zheng, C. H. Oh, H. Kim, J. I. Shim, D. S. Shin, and K. S. Kim, Nonradiative recombination mechanisms in InGaN/GaN-based light-emitting diodes investigated by temperature-dependent measurements, *Applied Physics Letters* **104** 151108 (2014)

[51] J. Iveland, M. Piccardo, L. Martinelli, J. Peretti, J. W. Choi, N. Young, S. Nakamura, J. S. Speck, and C. Weisbuch, Origin of electrons emitted into vacuum from InGaN light emitting diodes, *Applied Physics Letters* **105** 052103 (2014).

[52] D. P. Han, H. Kim, J. I. Shim, D. S. Shin, and K. S. Kim, Influence of carrier overflow on the forward-voltage characteristics of InGaN-based light-emitting diodes, *Applied Physics Letters* **105** 191114 (2014).

[53] J. Piprek, F. Römer, and B. Witzigmann, On the uncertainty of the Auger recombination coefficient extracted from InGaN/GaN light-emitting diode efficiency droop measurements. *Applied Physics Letters*, 106 101101 (2015).

[54] E. Jung, G. Hwang, J. Chung, O. Kwon, J. Han, Y. T. Moon, and T. Y. Seong, Investigating the origin of efficiency droop by profiling the temperature across the multi-quantum well of an operating light-emitting diode, *Applied Physics Letters* **106** 041114 (2015)

[55] X. Meng, L. Wang, Z. Hao, Y. Luo, C. Sun, Y. Han, B. Xiong, J. Wang, and H. Li, Study on efficiency droop

in InGaN/GaN light-emitting diodes based on differential carrier lifetime analysis, *Applied Physics Letters* **108** 013501 (2016).

[56] T. Kim, T. Y. Seong, and O. Kwon, Investigating the origin of efficiency droop by profiling the voltage across the multi-quantum well of an operating light-emitting diode. *Applied Physics Letters*, **108** 231101 (2016).

[57] H. Masui, S. Nakamura, S. P. DenBaars, and U. K. Mishra, Nonpolar and semipolar III-nitride light-emitting diodes: Achievements and challenges, *IEEE Transactions on Electron Devices* **57** 88 (2010).

[58] S. C. Ling, T. C. Lu, S. P. Chang, J. R. Chen, H. C. Kuo, and S. C. Wang, Low efficiency droop in blue-green m-plane InGaN/GaN light emitting diodes. *Applied Physics Letters* **96** 231101 (2010).

[59] Y. Zhao, S. Tanaka, C. C. Pan, K. Fujito, D. Feezell, J. S. Speck, S. P. DenBaars, and S. Nakamura, High-power blue-violet semipolar ($20\bar{2}\bar{1}$) InGaN/GaN light-emitting diodes with low efficiency droop at 200 A/cm$^2$. *Applied physics express* **4** 082104 (2011).

[60] C. C. Pan, S. Tanaka, F. Wu, Y. Zhao, J. Speck, S. Nakamura, S. P. DenBaars, and D. Feezell, High-power, low-efficiency-droop semipolar (2021) single-quantum-well blue light-emitting diodes, *Applied Physics Express* **5** 6 (2012).

[61] D. F. Feezell, J. S. Speck, S. P. DenBaars, and S. Nakamura, Semipolar InGaN/GaN light-emitting diodes for high-efficiency solid-state lighting, *Journal of Display Technology* **9** 190 (2013).

[62] D. L. Becerra, Y. Zhao, S. H. Oh, C. D. Pynn, K. Fujito, S. P. DenBaars, and S. Nakamura, High-power low-droop violet semipolar ($30\bar{3}\bar{1}$) InGaN/GaN light-emitting diodes with thick active layer design. *Applied Physics Letters* **105** 171106 (2014).

[63] M. J. Cich, R. I. Aldaz, A. Chakraborty, A. David, M. J. Grundmann, A. Tyagi, M. Zhang, F. M. Steranka, and M. R. Krames, Bulk GaN based violet light-emitting diodes with high efficiency at very high current density. *Applied Physics Letters* **101** 223509 (2012).

[64] C. A. Hurni, A. David, M. J. Cich, R. I. Aldaz, B. Ellis, K. Huang, A. Tyagi, R. A. DeLille, M. D. Craven, F. M. Steranka, and M. R. Krames, Bulk GaN flip-chip violet light-emitting diodes with optimized efficiency for high-power operation. *Applied Physics Letters* **106** 031101 (2015)

[65] A. Kaneta, M. Funato, and Y. Kawakami, Nanoscopic recombination processes in InGaN/GaN quantum wells emitting violet, blue, and green spectra, *Physical Review B* **78** 125317 (2008)

[66] S. F. Chichibu, A. Uedono, T. Onuma, B. A. Haskell, A. Chakraborty, T. Koyama, P. T. Fini, S. Keller, S. P. DenBaars, J. S. Speck, U. K. Mishira, S. Nakamura, S. Yamaguchi, S. Kamiyama, H. Amano, I. Akasaki, J. Han, and T. Sota, Origin of defect-insensitive emission probability in In-containing (Al, In, Ga)N alloy semiconductors, *Nature Materials* **5** 810 (2006)

[67] C. H. Yen, W. C. Lai, Y. Y. Yang, C. K. Wang, T. K. Ko, S. J. Hon, and S. J. Chang, GaN-based light-emitting diode with sputtered AlN nucleation layer. *IEEE Photonics Technology Letters* **24** 294 (2012).

[68] A. David and M. J. Grundmann, Influence of polarization fields on carrier lifetime and recombination rates in InGaN-based light-emitting diodes, *Applied Physics Letters* **97** 033501 (2010).

[69] A. E. Romanov, T. J. Baker, S. Nakamura, and J. S. Speck, Strain-induced polarization in wurtzite III-nitride semipolar layers, *Journal of Applied Physics*, **100** 023522 (2006).

[70] S. Nakamura, The roles of structural imperfections in InGaN-based blue light-emitting diodes and laser diodes, *Science*, **281** 956 (1998).

[71] Y. Kanitani, S. Tanaka, S. Tomiya, T. Ohkubo, and K. Hono, Atom probe tomography of compositional fluctuation in GaInN layers, *Japanese Journal of Applied Physics*, **55** 05FM04 (2016).

[72] R. A. Arif, Y. K. Ee, and N. Tansu, Polarization engineering via staggered InGaN quantum wells for radiative efficiency enhancement of light emitting diodes. *Applied Physics Letters*, **91** 091110 (2007).

**Acknowledgments:** The authors are highly indebted to the National Key Research and Development Program of China (Grant No. 2016YFB0400102), National Basic Research Program of China (Grant Nos. 2012CB3155605, 2013CB632804, 2014CB340002 and 2015CB351900), the National Natural Science Foundation of China (Grant Nos. 61574082, 61210014, 61321004, 61307024, and 51561165012), the High Technology Research and Development Program of China (Grant No. 2015AA017101), Tsinghua University Initiative Scientific Research Program (Grant No. 2013023Z09N, 2015THZ02-3), the Open Fund of the State Key Laboratory on Integrated Optoelectronics (Grant No. IOSKL2015KF10), the CAEP Microsystem and THz Science and Technology Foundation (Grant No.CAEPMT201505), Science Challenge Project (Grant No. JCKY2016212A503) and the Guangdong Province Science and Technology Program (Grant No. 2014B010121004).

**Author Contributions:** L. Wang designed the research and wrote the paper; L. Wang, X. Meng and J. Wang performed the experiments; L. Wang, X. Meng, J. Wang, Z. Hao, Y. Luo, C. Sun, Y. Han, B. Xiong, J. Wang and H. Li analyzed data; L. Wang supervised the project.

**Competing financial interests:** The authors declare no competing financial interests.

Table I. Fitting parameters of near-UV, blue, and green LEDs

| LEDs | $n_C$ (cm$^{-3}$) | $k$ | $A_L$ (s$^{-1}$) | $A_{NL}$ (s$^{-1}$) | $B_L$ (cm$^3$s$^{-1}$) | $B_{NL}$ (cm$^3$s$^{-1}$) |
|---|---|---|---|---|---|---|
| Near-UV | $1.35 \times 10^{17}$ | 2 | $1.25 \times 10^6$ | $1.9 \times 10^6$ | $18.5 \times 10^{-12}$ | $13 \times 10^{-12}$ |
| Blue | $1.5 \times 10^{17}$ | 2.8 | $0.25 \times 10^6$ | $1 \times 10^6$ | $14.8 \times 10^{-12}$ | $6.4 \times 10^{-12}$ |
| Green | $4.8 \times 10^{17}$ | 4.4 | $0.17 \times 10^6$ | $1.5 \times 10^6$ | $3 \times 10^{-12}$ | $2.6 \times 10^{-12}$ |

Figure captions

Fig. 1 (a) IQE and (b) carrier lifetimes depending on injection current density. Dot: experimental data; dash line: fitting curve.

Fig. 2 (a) CIE, (b) RRE, and (c) carrier concentration depending on injection current density.

Fig. 3 (a) IQE, (b) carrier lifetime, (c) CIE, and (d) RRE depending on carrier concentration.

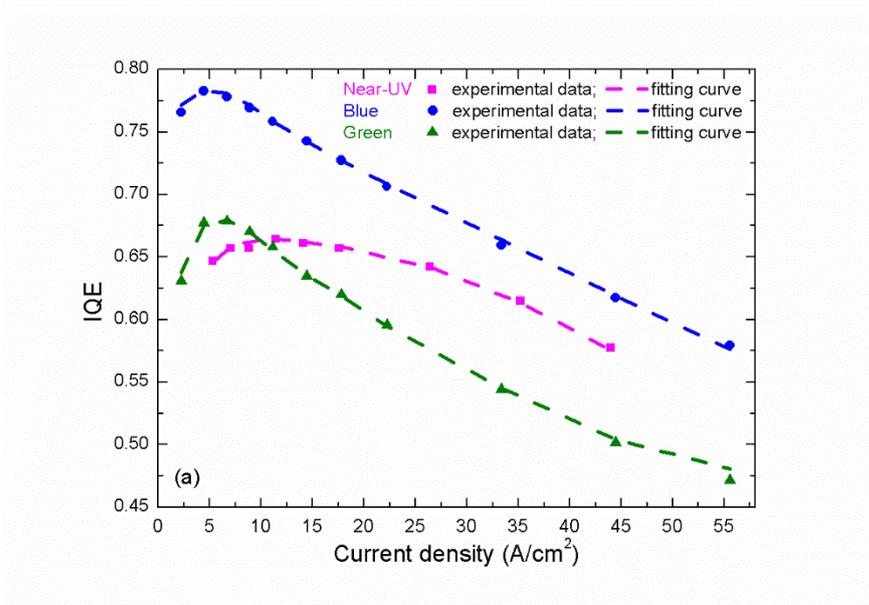

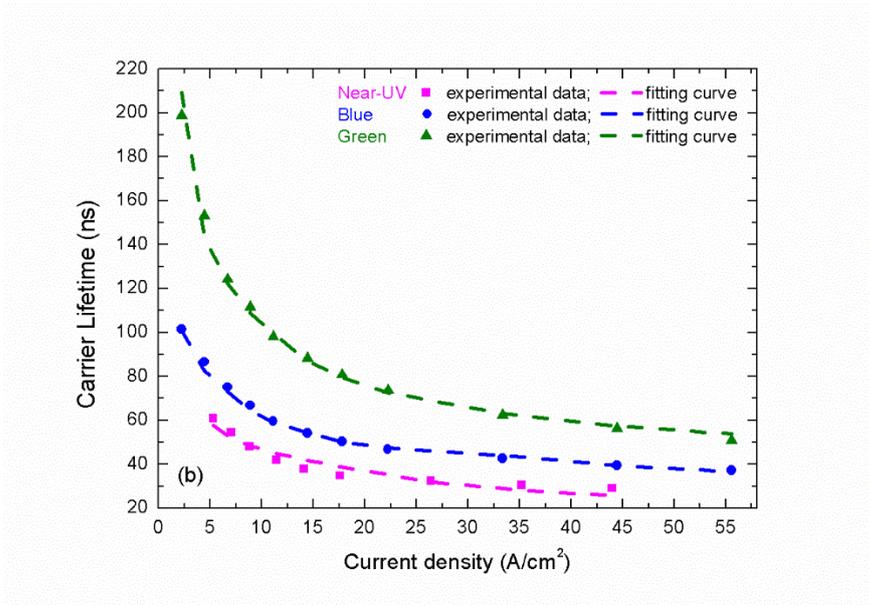

Fig. 1 L. Wang, *et al.*

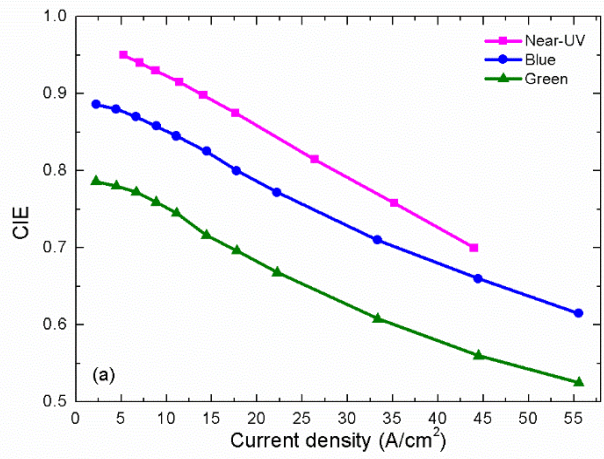

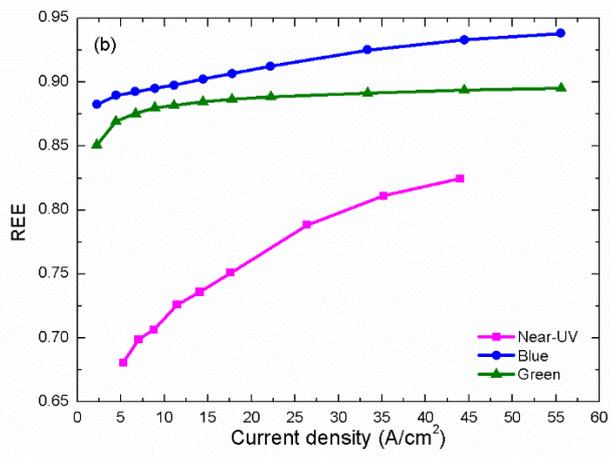

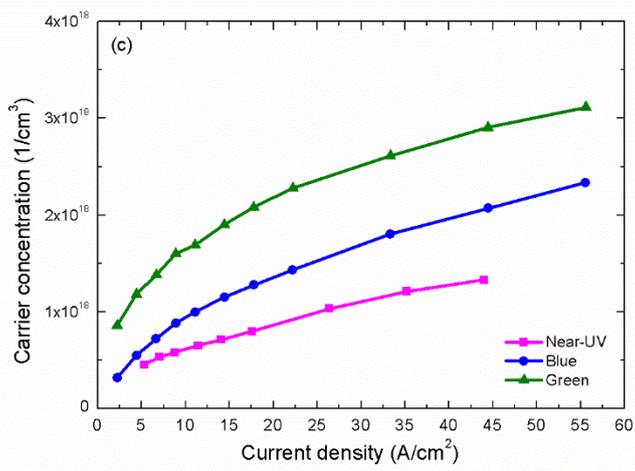

Fig. 2 L. Wang, *et al*.

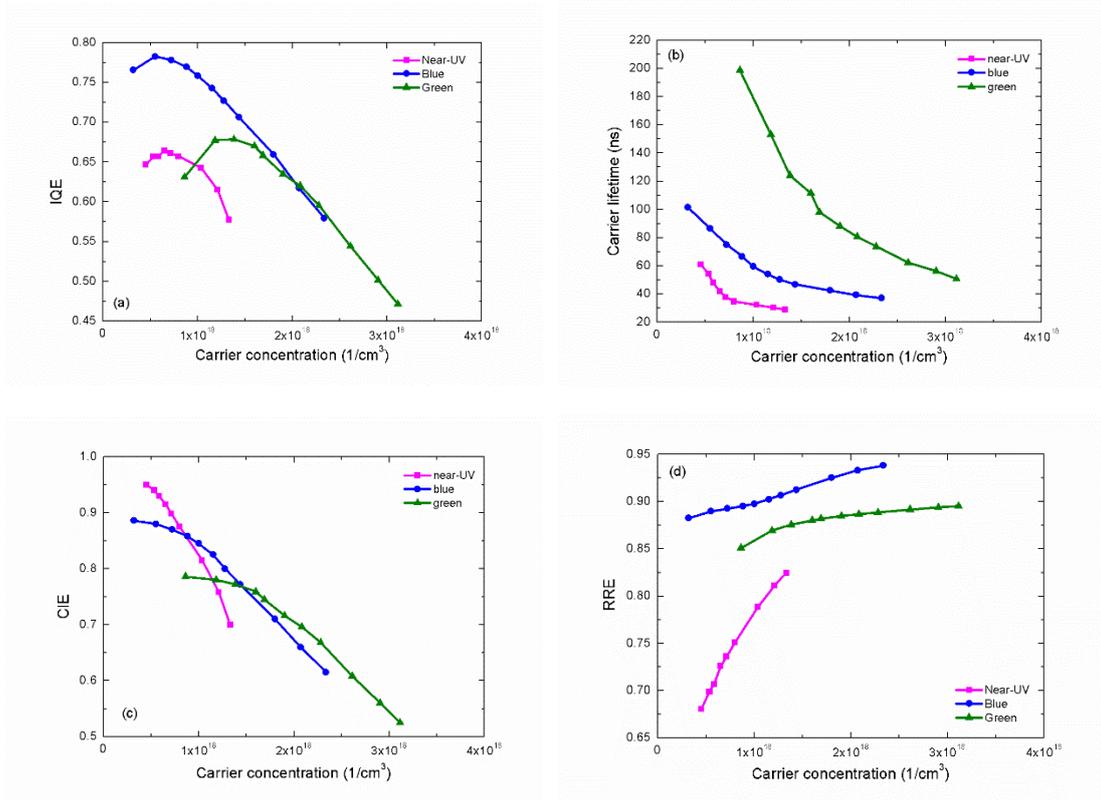

Fig. 3 L. Wang, *et al*.